# Low-complexity CV-QKD system with optical pilot-tone local oscillator synchronization

SAMAEL SARMIENTO*, JEISON TABARES, AND SEBASTIAN ETCHEVERRY

*LuxQuanta Technologies S.L., Mediterranean Technology Park. Carrer d'Esteve Terradas, 1, Office 206,08860, Castelldefels (Barcelona), Spain*
**samael.sarmiento@luxquanta.com*

**Abstract:** This study presents a comprehensive simulation-based analysis of optical pilot tone generation for local oscillator synchronization in continuous-variable quantum key distribution (CV-QKD) systems using Gaussian-modulated (GM) coherent states. We compare system performance when the pilot tone is generated either electrically or optically, examining the effects of key parameters such as pilot tone power, laser linewidth, digital-to-analog converter resolution, and link distance. The results provide insights into pilot tone generation strategies that enhance robustness while reducing the implementation complexity of GM CV-QKD systems.

## 1. Introduction

Continuous-variable quantum key distribution (CV-QKD) enables two parties to establish a shared secret key for secure communication [1,2]. Unlike its discrete-variable counterpart, which relies on individual photons and discrete states such as polarization or relative phase, CV-QKD exploits the continuous-variable properties of the amplitude and phase of electromagnetic lightwaves. In particular, coherent states carrying random information for key generation are prepared using electro-optical modulators, transmitted through an insecure channel (typically optical fiber), and measured by a coherent receiver employing either phase-diverse homodyne detection (PDHomD) [3–5] or radio-frequency heterodyne detection (RFHetD) [6–8], both aided by a local oscillator (LO) laser. Because CV-QKD leverages technologies already used in classical coherent telecommunications, it is well suited for integration into existing optical networks.

In PDHomD, the received signal and LO are split into two branches and measured using two balanced photodetectors (BPDs). By applying a $\pi/2$ phase shift to one branch, both quadratures (I and Q) of the received signal can be retrieved simultaneously [9]. In contrast, RFHetD sets the LO to a frequency $\omega_S - \omega_{IF}$ outside the signal band centered on $\omega_S$. Photodetection then downconverts the received signal to an intermediate frequency $\omega_{IF}$, preserving both quadratures, which can subsequently be recovered through RF downconversion in the electrical domain [9]. Thus, RFHetD requires only one symmetric beam splitter (BS) and a single BPD, whereas PDHomD requires two BPDs and four BSs, which increases optical complexity and doubles optical insertion losses. However, PDHomD naturally allocates signals to baseband after photodetection, leading to more efficient bandwidth utilization. By contrast, RFHetD requires BPDs with sufficiently large bandwidth to capture the full signal spectrum centered at $\omega_{IF}$. Receiver losses and noise directly impact the achievable key rate. Consequently, the reduced loss and noise inherent to RFHetD generally provide higher key rates and longer transmission distances, offsetting the need for increased detector bandwidth [6,10].

To maintain synchronization between the transmitter and LO lasers, early CV-QKD systems typically derived both the quantum and LO signals from the same laser and transmitted them together through the channel [11]. This approach, however, introduces both security vulnerabilities and practical limitations [12–14]. An alternative strategy is the use of a frequency-multiplexed pilot tone for optical phase synchronization [15]. Instead of transmitting the LO directly, the quantum signal is sent alongside a pilot tone, enabling the LO to be generated locally at the receiver. Although the pilot tone requires significantly less power than the LO, it must still be sufficiently strong to maintain synchronization under challenging conditions such as noisy, long-distance channels. In practice, the pilot tone often needs to be at least two orders of magnitude stronger than the quantum signal [6–8].

The large power disparity between the pilot tone and quantum signal frequently introduces severe crosstalk and distortion in Gaussian-modulated (GM) CV-QKD systems employing pilot-tone-based LO synchronization. A major source of distortion arises from the limited resolution of digital-to-analog and analog-to-digital converters (DACs/ADCs), which degrades modulation and detection quality, effectively increasing excess noise [6–8]. To mitigate this issue, prior studies have explored the use of a low-power pilot tone combined with Kalman- or window-based filters to reduce phase estimation errors in LO synchronization [7,16–19]. Another approach is to generate the pilot tone optically rather than electrically. Removing the pilot tone from the electrical signal input to the DAC reduces DAC-induced distortion as well as overall system complexity. Although the use of an optical pilot tone for LO synchronization in CV-QKD systems has been demonstrated [15,20,21], its advantages remain insufficiently explored.

In this study, we conduct a comprehensive simulation analysis to investigate the generation and use of an optical pilot tone for LO synchronization in GM CV-QKD systems employing RFHetD. Specifically, we analyze and compare system performance in terms of excess noise and secret key rate under varying system parameters (including pilot tone power, laser linewidth, DAC resolution, and link distance) when the pilot tone is generated either electrically or optically. The results indicate that, in long-distance channels requiring strong pilot tones, the optical pilot-tone approach achieves performance comparable to that of high-resolution DACs, even when using a low-resolution DAC (as low as 4 bits per quadrature), whereas the electrical pilot-tone approach proves ineffective. Furthermore, the optical pilot tone not only substantially reduces the implementation complexity of GM CV-QKD with RFHetD architectures but also enhances robustness by mitigating signal distortion, thereby reducing the risk of side-channel attacks.

The remainder of this paper is organized as follows. Section 2 describes the simulation framework, including pilot-tone generation and reception in both the electrical and optical domains, the simulation setup, the digital signal processing (DSP) chain, and parameter estimation using a multiprocessing technique. Section 3 presents the simulation results, and Section 4 concludes with the main findings.

## 2. Simulation framework

### 2.1 Electrical versus optical pilot-tone generation and reception

In low-complexity CV-QKD systems employing RFHetD, a pilot tone at a frequency distinct from that of the quantum signal is typically used to synchronize the transmitter (TX) laser and the receiver (RX) LO. In such systems, an optical in-phase-and-quadrature (IQ) modulator generates both the quantum signal and the pilot tone for transmission through the fiber channel. At the RX side, the incoming optical signal is combined with the LO during photodetection. In this subsection, we describe pilot-tone generation (both electrical and optical) and outline the corresponding reception technique required for proper LO synchronization.

The IQ modulator is built on a Mach–Zehnder (MZ) interferometer, with a nested MZ modulator (MZM) placed in each arm. One of the two arms includes an additional $\pi/2$ phase shift. Assuming push–pull operation and identical MZMs, the transfer function in terms of the electric field ratio ($E_{out}/E_{in}$) is

$$\frac{E_{out}}{E_{in}} = \frac{1}{2}\left\{\left(e^{j\frac{\pi(V_{RF1}-V_{Bias})}{2V_\pi}} + \gamma e^{-j\frac{\pi(V_{RF1}-V_{Bias})}{2V_\pi}}\right) + j\left(e^{j\frac{\pi(V_{RF2}-V_{Bias})}{2V_\pi}} + \gamma e^{-j\frac{\pi(V_{RF2}-V_{Bias})}{2V_\pi}}\right)\right\} \quad (1)$$

where $V_{RF1}$ and $V_{RF2}$ are the RF drive signals for the IQ modulator, $V_{Bias}$ is the bias voltage applied to each MZM, $V_\pi$ is the voltage required to induce a $\pi$ phase shift in the optical signal passing through the MZM, and $\gamma$ relates to the extinction ratio $\delta$ of the MZMs (in linear units) by $\gamma = (\sqrt{\delta} - 1)/(\sqrt{\delta} + 1)$. For high extinction ratios, the normalized output power per quadrature of the IQ modulator is $P_{out}^{I/Q} = cos^2(0.5\pi V_{Bias}/V_\pi)$. Thus, depending on the bias voltage $V_{Bias}$, two main operating points arise:

- **Null point**. This corresponds to the voltage required to achieve destructive interference where $P_{out}^{I/Q} = 0$. At this point, the optical carrier is largely suppressed, although ultimate suppression is limited by $\delta$. The pilot tone must therefore be added electrically, with $V_{RF1} = \cos(2\pi f_p t)$ and $V_{RF2} = \sin(2\pi f_p t)$, where $f_p$ is the pilot-tone frequency. Assuming an input field from the TX laser of $E_{in} = e^{j\Omega_{TX}t}$, the IQ modulator output is $E_{out}^{EP} \propto e^{j(\Omega_{TX}+2\pi f_p)t}$.
- **Quadrature point.** This occurs when the MZMs are biased at half their $V_\pi$, yielding $P_{out}^{I/Q} = 1/2$. Unlike the null point, the optical carrier is not suppressed. In the absence of RF drive signals, the output field is $E_{out}^{OP} \propto e^{j\Omega_{TX}t}$.

Figure 1 compares electrical and optical pilot-tone generation alongside the quantum signal. The power spectral density (PSD) of both the electrical drive signals and the resulting optical output at the IQ modulator is shown for the null and quadrature operating points. When modulating random data, the output field becomes $E_{out}^{EP} \propto e^{j(\Omega_{TX}+2\pi f_p)t} + a(t)e^{j(\Omega_{TX}+2\pi f_{uc}^{EP})t}$ for the electrical pilot (EP), and $E_{out}^{OP} \propto e^{j\Omega_{TX}t} + a(t)e^{j(\Omega_{TX}+2\pi f_{uc}^{OP})t}$ for the optical pilot (OP). Here, $a(t)$ is the quantum signal $(X + jP)$, typically upconverted to frequency $f_{uc}^{EP/OP}$ to avoid baseband distortion caused by AC-coupled electronics. Importantly, when the pilot tone is generated optically, the dynamic range of the DAC can be tailored to the quantum signal without degrading modulation quality (Figure 1). In contrast, electrical pilot tones require the DAC to accommodate both signals, reducing the effective number of bits available for the quantum signal. Moreover, because the optical carrier carries the same phase information as the electrical pilot, it can be used for LO synchronization.

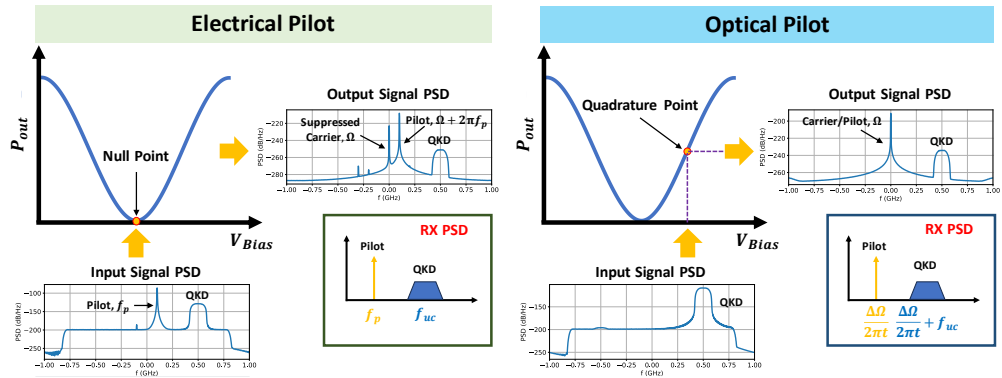

Fig. 1. Electrical vs. optical pilot-tone generation: power spectral density (PSD) of the electrical signals driving the IQ modulator and corresponding optical PSD at the modulator output. (Inset) PSD representation at the receiver output. A modulator extinction ratio of 35 dB, $V_\pi$ = 1 V, and single side-band modulation, were assumed.

After mixing with the LO field $E_{LO} = e^{j\Omega_{LO}t}$ and photodetection, the photodetected electrical and optical pilot currents are $I_{det}^{EP} \propto |a(t)|cos(\Delta\Omega t + 2\pi f_{uc}^{EP} t + \Delta\phi) + cos(\Delta\Omega t + 2\pi f_p t + \Delta\phi)$ and $I_{det}^{OP} \propto |a(t)|cos(\Delta\Omega t + 2\pi f_{uc}^{OP} t + \Delta\phi) + cos(\Delta\Omega t + \Delta\phi)$, respectively, where $\Delta\Omega = \Omega_{TX} - \Omega_{LO}$ and $\Delta\phi$ represent the frequency mismatch and phase noise from the lasers, respectively. For electrical pilots, $\Delta\Omega/2\pi t$ should be close to 0, whereas for optical pilots, it must exceed the bandwidth of phase noise caused by laser beating. The inset of Figure 1 shows the PSD of the receiver output after photodetection. At this stage, the pilot tone is filtered, and its phase noise, $\Delta\phi$, can be subtracted from the quantum signal phase, enabling LO synchronization.

Finally, another key distinction between the two approaches concerns modulator stabilization in practical systems. At the null point, stabilization typically requires a strong high-frequency dithering tone (on the kilohertz scale) because of the low output power, which increases complexity. At the quadrature point, by contrast, stabilization can be achieved directly through the output power, simplifying implementation.

### 2.2 Simulation setup, DSP, and parameter estimation

Figure 2 shows the optical simulation setup of a low-complexity CV-QKD system based on an RFHetD configuration, where the LO is generated locally at the RX [6–8]. A continuous-wave (CW) laser was fed into an IQ modulator, which prepared the coherent states of light according to the GM protocol. An $n_{DAC}$-bit DAC operating at the sampling frequency $F_s$ generated the RF signals ($V_{RF1}$ and $V_{RF2}$) that drove the IQ modulator. The working point of the IQ modulator (null or quadrature) was set by the bias voltage $V_{Bias}$. Once the coherent states were prepared, their power was reduced to the quantum level using a variable optical attenuator (VOA). An in-line power meter (PoM) estimated the modulation variance ($V_{mod}$) at the TX output. A 100 km optical fiber link was initially considered as the quantum channel, with distances up to 200 km also investigated. At the RX, the optical signal was combined with a CW laser serving as the LO using a 50:50 beam splitter (BS) and then detected with a BPD. The electrical output was digitized using a 12-bit ADC operating at $F_s$. Two optical switches (OSs) were included to calibrate electronic and shot noise in accordance with the GM protocol.

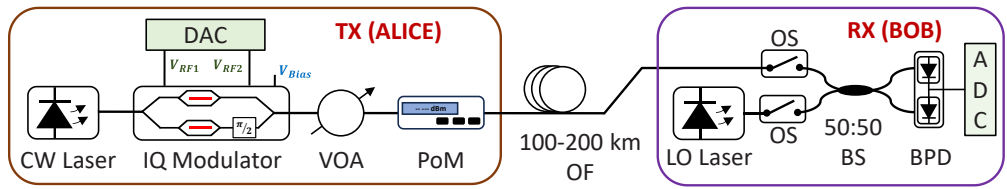

Fig. 2. Simulation setup. (Acronyms) CW laser: continuous-wave laser; IQ modulator: in-phase-and-quadrature modulator; DAC: digital-to-analog converter; VOA: variable optical attenuator; PoM: power meter; OF: optical fiber; LO laser: local oscillator laser; OS: optical switch; BS: beam splitter; BPD: balanced photodetector; ADC: analog-to-digital converter.

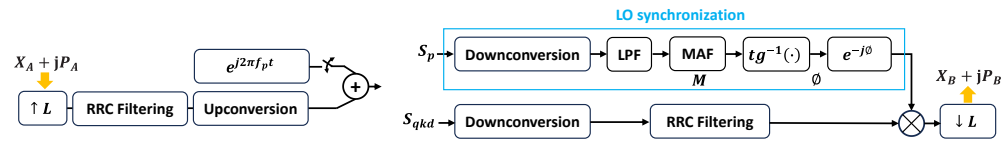

Fig. 3. Digital signal processing (DSP) chain. (Acronyms) RRC: root-raised-cosine filter; L: upsampling/downsampling factor, LPF: low-pass filter, MAF: moving average filter.

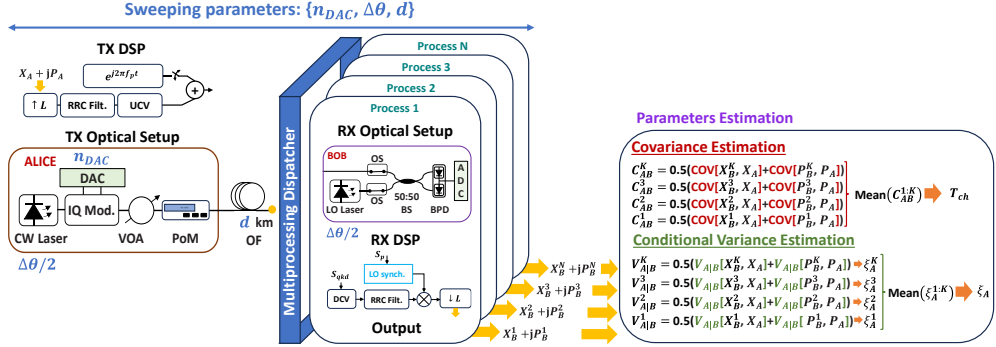

Fig. 4. Parameter estimation framework using a multiprocessing approach. $K$ copies of the optical signal are distributed across $N$ processors to perform optical reception, including shot and electronic noise calibration, as well as DSP operations to retrieve the quantum data. Parameter estimation is then performed. The simulator can easily be adapted to sweep different parameters such as DAC resolution ($n_{DAC}$), total spectral linewidth ($\Delta\theta = \Delta\theta_{TX} + \Delta\theta_{RX}$), and link distance ($d$).

Figure 3 presents the DSP chain. Two Gaussian-distributed symbol sets of size $N_{sym}$, $X_A$ and $P_A$, were upsampled to $F_s/R_s$ by inserting zeros, where $R_s$ denotes the symbol rate of the quantum signal. After upsampling, the complex signal was shaped using a root-raised-cosine (RRC) filter characterized by its roll-off factor, and then upconverted to $f_{uc}$ in a single-sideband configuration. Optionally, a digital pilot tone at $f_p$ could be added for synchronization of the TX and RX lasers. At the receiver, the DSP chain included LO synchronization via the pilot-tone technique, downconversion to baseband, RRC filtering, and downsampling. Synchronization accuracy was further improved by applying a moving average filter with $M$ samples to each quadrature of the pilot signal [18]. The resulting outputs were two symbol sets, $X_B$ and $P_B$, which correlate with $X_A$ and $P_A$, respectively.

Figure 4 illustrates the parameter estimation framework. Because statistical fluctuations are on the same order as the parameters to be estimated, a large dataset was accumulated so that the fluctuations average out and the estimates become more reliable. To manage the resulting computational load, a multiprocessing approach was employed. First, the Gaussian-distributed symbols were processed and sent to the DAC unit for optical signal generation using the IQ modulator. After transmission through the optical channel, $K$ copies of the optical signal were distributed across $N$ processors, with $K \gg N$ and each processor serving, on average, $K/N$ signals. Each processor performed optical reception, including shot and electronic noise calibration, as well as the DSP operations necessary to retrieve the quantum data (see Figure 3). Parameter estimation was carried out in the asymptotic regime, assuming reverse reconciliation. The modulation variance was calculated as $V_{mod} = 2\langle n \rangle$, where $\langle n \rangle$ is the mean photon number per symbol, defined as $P_{PoM}[(1+\rho)E_{ph}R_s]^{-1}$, with $P_{PoM}$ being the power measured by the inline PoM, $\rho$ the power ratio between the pilot tone and the quantum states, and $E_{ph}$ the photon energy at the working wavelength. The channel transmittance, $T_{ch}$, was estimated by averaging the covariance between transmitted and received data ($C_{AB}$) across the $K$ signals (Figure 4): $T_{ch} = (2/\eta)(\langle C_{AB}^{1:K}\rangle/V_{mod})^2$, with $\eta$ being the detection efficiency, calibrated beforehand. The excess noise referred to in the channel input was calculated in shot-noise units (SNU) as $\xi_A = \langle \xi_A^{1:K} \rangle = 2(\langle V_{A|B}^{1:K}\rangle - 1 - V_{en})/\eta T_{ch}$, where $V_{B|A}$ is the conditional variance estimator and $V_{en}$ is the electronic noise variance. $V_{B|A}$ was estimated as the variance of $X_B - \sqrt{0.5\eta T}X_A$ for the $X$ quadrature and $P_B - \sqrt{0.5\eta T}P_A$ for the $P$ quadrature. Finally, the secret key rate ($SKR$) in bits per second (bps) was calculated as $0.5(\beta I_{AB} - \chi_{BE})R_{eff}$, where $\beta$ is the reconciliation efficiency (typically 0.95 [11]), $I_{AB}$ is the mutual information between the transmitted (Alice) and received (Bob) data, $\chi_{BE}$ is the Holevo bound representing the maximum information leakage to an eavesdropper, and $R_{eff} = R_s/2$ is the effective symbol rate, with half of the data allocated to parameter estimation.

Table 1. Simulation Parameters

| Parameter | Value | Parameter | Value |
|---|---|---|---|
| Sampling freq. ($F_s$) | 2 GSa/s | PD responsivity | 1 A/W |
| Symbol rate ($R_s$) | 100 MHz | PD noise equivalent power | $7 \cdot 10^{-12}$ W/$\sqrt{Hz}$ |
| Upconversion freq. ($f_{uc}^{EP}$) in EP | 500 MHz | Transimpedance amplifier gain | 3500 V/A |
| Upconversion freq. ($f_{uc}^{OP}$) in OP | 400 MHz | PD bandwidth | 800 MHz |
| RRC num. symbols | 20 | ADC resolution | 12 bits |
| RRC roll-off factor | 0.65 | Optical fiber loss coefficient | 0.16 dB/km |
| MZM $V_\pi$ | 1 V | Number of QKD symbols ($N_{sym}$) | $10^7$ |
| MZM extinction ratio | 35 dB | Number of copies of the optical signal ($K$) | 1250 |
| IQ ratio imbalance | 0 dB | Number of processors ($N$) | 10 |

## 3. Analysis and results

Table 1 summarizes the key parameters used in the simulations. Figures 5(a) and 5(b) compare system performance in terms of the power ratio between the pilot and quantum signals ($\rho$) and the modulation variance ($V_{mod}$) as functions of DAC resolution ($n_{DAC}$), for both electrical and optical pilot-tone generation approaches. The analysis was performed at $V_{mod} = 2.5$ SNU, a value within the range where $SKR$ reaches its maximum and remains relatively stable before decaying at transmission distances of 100–200 km. Figure 5(a) shows that at higher $\rho$ values, the electrical pilot-tone technique requires DACs with higher resolution (larger $n_{DAC}$ values) to generate the target $\rho$, thus matching the performance of the optical pilot tone. As the estimation of $V_{mod}$ depends mainly on $\rho$ (see Section 2.2), a similar trend is observed in Figure 5(b). In both cases ($\rho$ and $V_{mod}$) the optical pilot-tone approach remains largely unaffected by DAC resolution. Figure 5(c) shows the optical signal PSD at the transmitter output for $n_{DAC} = 2$ and $n_{DAC} = 10$, with $\rho$ fixed at 34 dB. In the worst case, with $n_{DAC} = 2$, no spectral replicas are observed for the optical pilot tone, whereas the electrical pilot tone exhibits significant distortion. In this case, replicas of the electrical pilot tone overlap the spectral region occupied by the quantum signal, leading to an overestimation of $V_{mod}$. This explains the trends observed in Figures 5(a) and 5(b). Because the optical pilot-tone approach is less affected by distortion, it enables the use of low-cost, low-resolution DAC units while maintaining a low risk of potential information leakage that could be exploited in side-channel attacks.

To evaluate the benefits of the optical pilot tone with respect to DAC resolution, and to ensure a fair comparison between the two techniques in terms of $\xi_A$ and $SKR$, it is necessary to isolate the noise introduced by the LO synchronization procedure described in Section 2.2. For this purpose, the DAC resolution was set to a high value (e.g., 12 bits per quadrature). In this configuration, the analysis becomes independent of the pilot-tone generation technique, meaning that the noise floor originates solely from statistical effects rather than distortions induced by the system. Figure 6(a) presents the excess noise as a function of the frequency separation ($\Delta f$) between the pilot tone and the quantum signal, for different $\rho$ values. This analysis assumes a total spectral linewidth of the system ($\Delta\theta = \Delta\theta_{TX} + \Delta\theta_{RX}$) equal to zero, with LO synchronization disabled and $n_{DAC} = 12$. Ideally, since $\Delta\theta = 0$, there is no potential crosstalk from the pilot to the quantum signal, even for a small $\Delta f$ of 100 MHz and a large $\rho$ value of 34 dB. As shown in Figure 6(a), a noise floor of approximately 2.5 mSNU is observed, which, as expected, is independent of both $\Delta f$ and $\rho$. This result confirms the consistency of the simulations. The noise floor can be reduced by increasing the number of simulated symbols, though at the expense of longer simulation times. As a compromise, the total ensemble length was fixed at $KN_{sym} = 10^{10}$ symbols (see Table 1). Figure 6(b) repeats the simulation analysis of Figure 6(a), but with $\Delta\theta$ set at 200 Hz, LO synchronization enabled, and no phase averaging applied ($M$ = 0). Under these conditions, system performance becomes highly dependent on $\rho$ due to the influence of additive noise (shot and electronic noise) on the pilot measurement. As a result, lower excess noise levels are achieved with stronger pilot tones. Figure 6(b) also shows that marginal $SKR$ values occur when $\rho$ falls below 23 dB. Enabling averaging in the LO synchronization procedure improves performance, as shown in Figure 6(c). For example, at $\rho$ = 21 dB, the noise level decreases from 105 mSNU (Figure 6(b)) to 20 mSNU with $M$ = 500 samples (equivalent to 25 symbol periods of the quantum signal, and considerably smaller than the coherence time of the simulated spectral linewidth), and further down to 5 mSNU with $M$ = 5000 samples (250 symbol periods, also smaller than the coherence time). Since averaging a large number of samples significantly increases simulation time, $M$ was fixed at 2000 samples, which provides performance comparable to $M$ = 5000 while maintaining a reasonable runtime. Figure 6(d) shows the resulting performance with $M$ = 2000 for different $\rho$ values. With this configuration, the minimum $\rho$ required to achieve positive $SKR$ decreases from approximately 23 dB to 10 dB. Figure 6(d) also shows that for $\rho$ values above 21 dB, $\xi_A$ converges to 5 mSNU, the limit imposed by the chosen averaging window length $M$. Finally, Figure 7 examines the maximum tolerable value of $\Delta\theta$ that reduces $SKR$ to zero for different $\rho$ values, and fixed $M = 2000$. The key outcome is that, for any value of $\rho$, the $SKR$ remains constant up to $\Delta\theta = 2$ kHz. Beyond this point, it begins to degrade significantly, establishing the bound for the maximum laser linewidth at a given $\rho$.

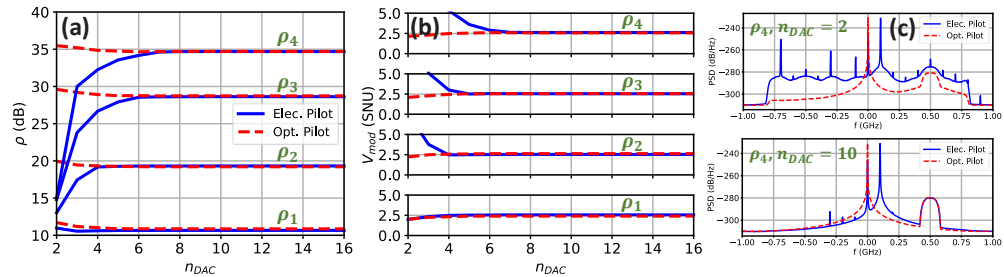

Fig. 5. $\rho$ (a) and $V_{mod}$ (b) versus DAC resolution ($n_{DAC}$). Optical signal PSD at the transmitter output for $n_{DAC} = 2$ and $n_{DAC} = 10$.

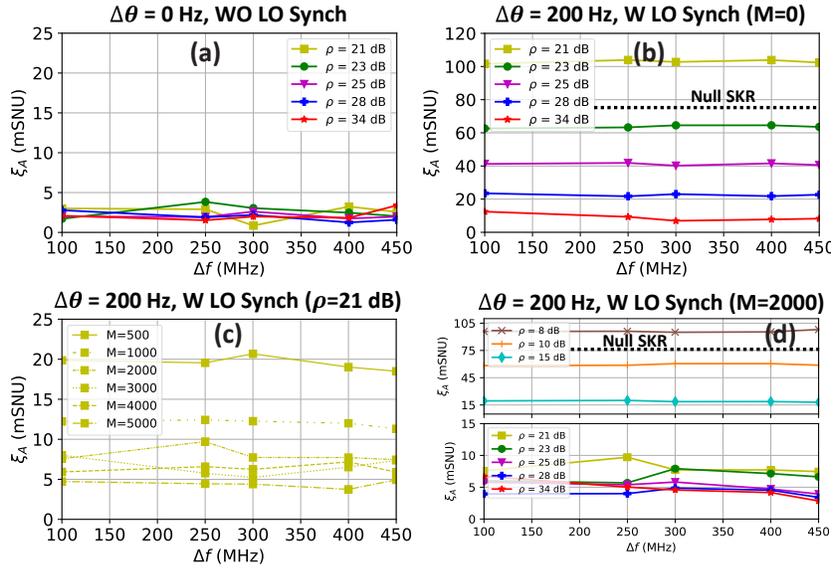

Fig. 6. Excess noise ($\xi_A$) performance when: (a) total spectral linewidth ($\Delta\theta$) is zero and LO synchronization is disabled, (b) $\Delta\theta$ = 200 Hz and LO synchronization is enabled for M = 0, (c) $\rho$ = 21 dB for different phase-averaging window $M$, and (d) $M$ = 2000 for different $\rho$ values.

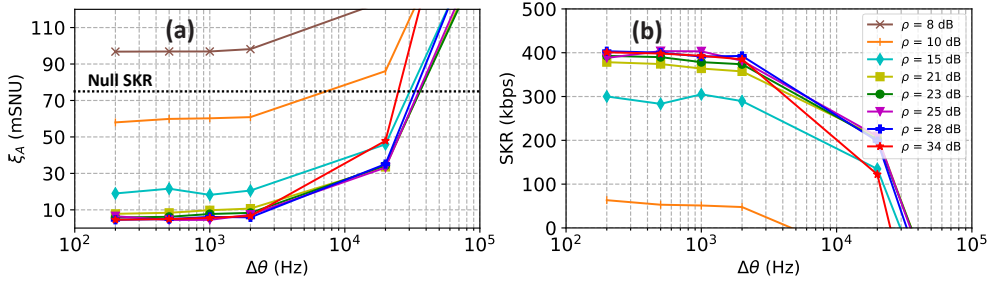

Fig. 7. System performance evaluation with $M = 2000$ as a function of total spectral linewidth ($\Delta\theta$) of the system for different $\rho$ values: (a) excess noise ($\xi_A$) and (b) secret key rate ($SKR$).

Figure 8 compares the electrical and optical pilot-tone generation approaches in terms of $\xi_A$ and $SKR$ as a function of $n_{DAC}$ for different values of $\rho$. For $\rho = 34$ dB, the electrical pilot-tone scheme begins to suffer from significant distortion when $n_{DAC}$ drops below 8 bits, resulting in zero $SKR$ when $n_{DAC}$ falls below 7 bits. By contrast, the optical pilot tone shows noticeable performance degradation only when $n_{DAC}$ drops below 3 bits. Figure 8 also shows that reducing $\rho$ relaxes the DAC resolution requirements for the electrical pilot-tone approach to achieve positive $SKR$. This trend suggests that at lower $\rho$ values, the system becomes less sensitive to the quantization noise introduced by the DAC. For example, at $\rho = 21$ dB, the electrical pilot tone requires a minimum resolution of 5 bits to keep the degradation in $\xi_A$ within acceptable limits for non-zero $SKR$, whereas the optical pilot tone achieves the same noise level with only 3 bits per quadrature, demonstrating superior resilience to quantization effects. However, decreasing $\rho$ below 21 dB results in performance degradation for both approaches, highlighting a trade-off between $\rho$ and DAC resolution for a given fiber channel distance. Beyond this point, lowering $\rho$ is no longer beneficial and leads to suboptimal key rates. The advantage of the optical pilot-tone approach becomes increasingly evident at lower $n_{DAC}$ values, particularly across a wide range of high $\rho$ values. For 21 dB $\leq \rho \leq$ 34 dB, the performance in $\xi_A$ and $SKR$ remains comparable (Figures 8(b) and 8(d)). This is especially relevant for extending transmission distance, which is often constrained by poor phase recovery accuracy due to hardware limitations such as DAC resolution.

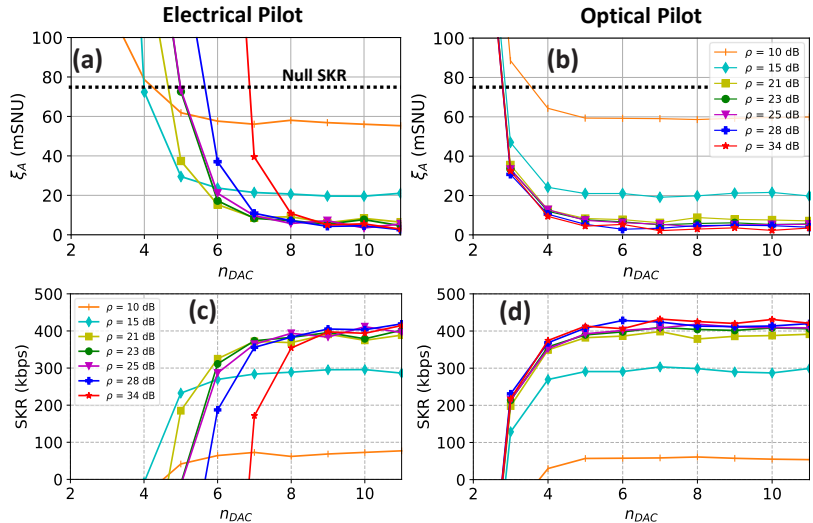

Fig. 8. Performance comparison between electrical and optical pilot tone approaches in terms of $\xi_A$ (a, b) and $SKR$ (c, d) as a function of $n_{DAC}$ for different $\rho$ values.

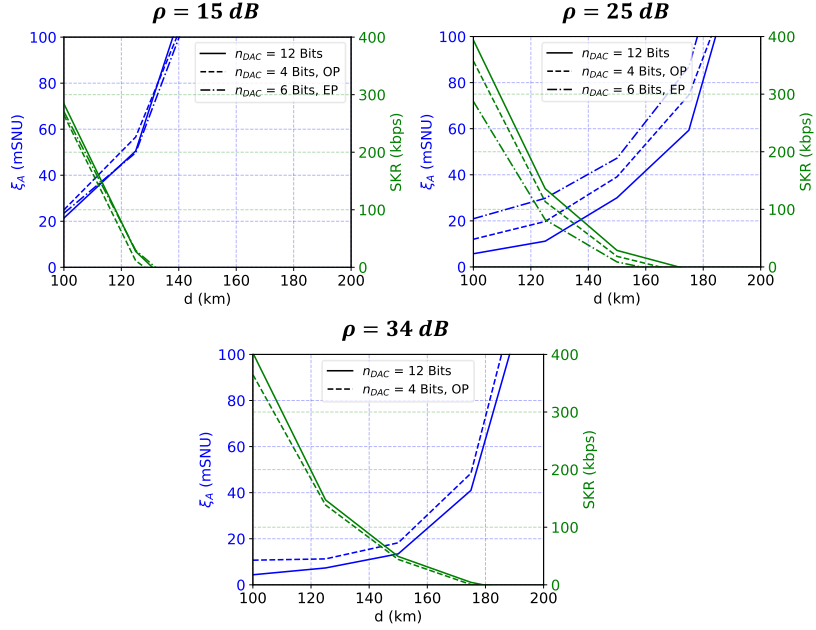

Fig. 9. System performance versus link distance d for $\rho = \{15, 25, 34\}$ dB and ADC resolutions of 4, 6, and 12 bits. Optical pilot (OP) and electrical pilot (EP) tone approaches are applied for 4 and 6 bits, respectively.

Figure 9 presents system performance as a function of link distance ($d$) for $\rho = \{15, 25, 34\}$ dB under three different DAC configurations, $n_{DAC} = \{4, 6, 12\}$ bits. For $n_{DAC} = 4$, the optical pilot-tone approach was employed; for $n_{DAC} = 6$, the electrical pilot tone was used. These configurations were selected based on the analysis in Figures 8(a) and 8(b), assuming cost-effective DACs available in the market. Figure 9 shows that for $\rho = 15$ dB, all three configurations achieve similar performance, with a maximum link distance of 130 km. Increasing $\rho$ to 25 dB extends the maximum reach to 170 km, where the optical pilot tone with a 4-bit DAC outperforms the electrical pilot tone with 6 bits and approaches the performance of a 12-bit DAC. At $\rho = 34$ dB, the reach extends further to 180 km. In this case, only the 4-bit optical pilot tone reaches the performance of the 12-bit DAC, while the 6-bit electrical pilot tone fails to generate positive $SKR$. These results confirm that the optical pilot tone tolerates lower-resolution DACs without significant performance loss, making it a scalable and practical solution for quantum key distribution over extended fiber links with reduced implementation complexity.

## 4. Conclusions

This study presented a detailed analysis of the optical pilot-tone technique for local oscillator synchronization in GM CV-QKD systems employing radio-frequency heterodyne detection. To benchmark its performance, we compared it with systems using conventional electrically generated pilot tones. Both approaches were evaluated across a range of realistic system parameters, including pilot-tone power, laser linewidth, DAC resolution, and transmission distance. The results show that the optical pilot-tone approach significantly enhances system robustness, particularly under hardware constraints such as limited DAC resolution and extended transmission distances. In particular, using an optical pilot tone with 4-bit DACs, a distance of 180 km was achieved, whereas the electrical pilot tone required at least 8-bit DACs to reach the same performance. These findings provide practical guidance for designing GM CV-QKD systems with improved performance and reduced implementation complexity, paving the way for more scalable and hardware-efficient metropolitan CV-QKD networks.

**Funding.** This work was partly supported by the European Union through the Digital Europe Work Programme 2021-2022 [DIGITAL-2021-QCI-01-INDUSTRIAL/QuarterProject/101091588], the European Innovation Council's Horizon Europe EIC Accelerator Programme under project MIQRO [101161539], and the Spanish Ministry (Grants PTQ2021-012121 and PTQ2021-012123), funded by [MCIN/AEI/10.13039/50110001103].

**Disclosures.** The authors declare no conflicts of interest.

**Data availability.** The data underlying the results presented in this paper are not publicly available at this time but may be obtained from the authors upon reasonable request.